\documentclass[superscriptaddress,showpacs,preprintnumbers,amsmath,amssymb,twocolumn,prx]{revtex4-1}
\usepackage{graphicx}
\usepackage{verbatim} 
\usepackage[dvips]{color} 
\usepackage{epstopdf}

\usepackage{dcolumn}
\usepackage{bm}
\usepackage{upgreek}

\begin{document}

\preprint{APS/123-QED}

\title{Long-range p-d exchange interaction in a ferromagnet-semiconductor Co/CdMgTe/CdTe quantum well hybrid structure}

\author{I.~A.~Akimov}
 	\affiliation{Experimentelle Physik 2, Technische Universit\"at Dortmund, 44221 Dortmund, Germany}
 	\affiliation{Ioffe Physical-Technical Institute, Russian Academy of Sciences, 194021 St. Petersburg, Russia}
\author{M.~Salewski}
 	\affiliation{Experimentelle Physik 2, Technische Universit\"at Dortmund, 44221 Dortmund, Germany}
\author{I.~V.~Kalitukha}
 	\affiliation{Ioffe Physical-Technical Institute, Russian Academy of Sciences, 194021 St. Petersburg, Russia}
\author{S.~V.~Poltavtsev}
 	\affiliation{Experimentelle Physik 2, Technische Universit\"at Dortmund, 44221 Dortmund, Germany}
 	\affiliation{Spin Optics Laboratory, St. Petersburg State University, 198504 St. Petersburg, Russia}
\author{J.~Debus}
\author{D.~Kudlacik}
    \affiliation{Experimentelle Physik 2, Technische Universit\"at Dortmund, 44221 Dortmund, Germany}
\author{V.~F.~Sapega}
    \affiliation{Ioffe Physical-Technical Institute, Russian Academy of Sciences, 194021 St. Petersburg, Russia}
\author{N.~E.~Kopteva}
 	\affiliation{Spin Optics Laboratory, St. Petersburg State University, 198504 St. Petersburg, Russia}
\author{E.~Kirstein}
\author{E.~A.~Zhukov}
    \affiliation{Experimentelle Physik 2, Technische Universit\"at Dortmund, 44221 Dortmund, Germany}
\author{D.~R.~Yakovlev}
 	\affiliation{Experimentelle Physik 2, Technische Universit\"at Dortmund, 44221 Dortmund, Germany}
 	\affiliation{Ioffe Physical-Technical Institute, Russian Academy of Sciences, 194021 St. Petersburg, Russia}
\author{G.~Karczewski}
 	\affiliation{Institute of Physics, Polish Academy of Sciences, PL-02668 Warsaw, Poland}
\author{M.~Wiater}
 	\affiliation{Institute of Physics, Polish Academy of Sciences, PL-02668 Warsaw, Poland}
\author{T.~Wojtowicz}
 	\affiliation{Institute of Physics, Polish Academy of Sciences, PL-02668 Warsaw, Poland}
    \affiliation{International Research Centre MagTop, PL-02668 Warsaw, Poland}
\author{V.~L.~Korenev}
\author{Yu.~G.~Kusrayev}
  	\affiliation{Ioffe Physical-Technical Institute, Russian Academy of Sciences, 194021 St. Petersburg, Russia}
\author{M.~Bayer}
 	\affiliation{Experimentelle Physik 2, Technische Universit\"at Dortmund, 44221 Dortmund, Germany}
 	\affiliation{Ioffe Physical-Technical Institute, Russian Academy of Sciences, 194021 St. Petersburg, Russia}

\date{\today}

\begin{abstract}
The exchange interaction between magnetic ions and charge carriers in semiconductors is considered as prime tool for spin control. Here, we solve a long-standing problem by uniquely determining the magnitude of the long-range $p-d$ exchange interaction in a ferromagnet-semiconductor (FM-SC) hybrid structure where a 10~nm thick CdTe quantum well is separated from the FM Co layer by a CdMgTe barrier with a thickness on the order of 10~nm. The exchange interaction is manifested by the spin splitting of acceptor bound holes in the effective magnetic field induced by the FM. The exchange splitting is directly evaluated using spin-flip Raman scattering by analyzing the dependence of the Stokes shift $\Delta_S$ on the external magnetic field $B$. We show that in strong magnetic field $\Delta_S$ is a linear function of $B$ with an offset of $\Delta_{pd} = 50-100~\mu$eV  at zero field from the FM induced effective exchange field. On the other hand, the $s-d$ exchange interaction between conduction band electrons and FM, as well as the $p-d$ contribution for free valence band holes, are negligible. The results are well described by the model of indirect exchange interaction between acceptor bound holes in the CdTe quantum well and the FM layer mediated by elliptically polarized phonons in the hybrid structure.
\end{abstract}

\maketitle

\section{Introduction}

The integration of magnetism into semiconductor electronics would initiate a new generation of computers based on advanced functional elements where the magnetic memory and electronic data processor are located on a single chip~\cite{Dietl-2010, Zutic-2004, Korenev-UFN-2005, Johnson-Spinbook}. One approach in this direction is based on hybrid systems where a thin ferromagnetic film is placed on top of a semiconductor. In such a system one expects to detect emergent functional properties which appear and benefit from bringing the primary constituents together, i.e. the magnetism as in ferromagnets (FM) with the optical and electrical tunability as in semiconductors (SC) ~\cite{Dzhioev-FTT-1995,Kawakami-2001,Petrou-2003,Crooker-2005,Crowell-2007,Jonker-2007,Ciorga-2009,Song-2011}. For that purpose it is mandatory to establish a strong exchange interaction between the charge carriers in the SC and the magnetic ions in the FM. Control of the concentration of the charge carriers and the penetration of their wavefunction into the FM layer should consequently change the magnitude of the exchange coupling between FM and SC~\cite{Korenev-2003}. As a result of the coupling, the following interdependencies are established: spin polarization of charge carriers in the SC by the magnetized FM layer and inverse action of the spin polarized carriers to control the FM magnetization. Previously, it was demonstrated that the stray fields of a FM layer influence the spin polarization of conduction band electrons in bulk GaAs~\cite{Dzhioev-FTT-1995, Jansen-2011} and diluted magnetic semiconductors~\cite{Crowell-1997, Henne-2007}. In turn, illumination of a GaAs SC changed the coercive force of a nickel-based interfacial FM layer (photocoercivity), which was attributed to optical control of the exchange coupling at the interface between FM and SC~\cite{Dzhioev-FTT-1995}.

A novel type of hybrid structure with a thickness of a few tens of nanometers only is obtained by combining a FM layer and a SC quantum well (QW) that are located in close proximity of each other, separated by a SC barrier with a few nanometer thickness~\cite{Crowell-1997, Henne-2007, Myers-2004, Aronzon-2009, Zaitsev-2010, NC-Korenev-2012, NP-Korenev-2016}. Such structures with a well-defined profile along the growth axis can be fabricated with monolayer precision. The stray fields from the FM layer are weak so that they contribute significantly to the carrier spin polarization only in combination with a magnetic SC QW \cite{Crowell-1997, Henne-2007}. For non-magnetic SCs, the contributing mechanisms are a direct exchange interaction generating an equilibrium spin polarization~\cite{Korenev-2003,Myers-2004,Zaitsev-2010,Aronzon-2009} and a spin dependent tunneling into the FM layer~\cite{NC-Korenev-2012,Rozhansky-2015}. In Ref.~\onlinecite{NC-Korenev-2012} it was demonstrated that in hybrid structures based on combining a GaMnAs FM with a InGaAs QW the conduction band electrons in the QW are spin polarized due to spin-dependent capture into the FM layer. Another mechanism leading to an equilibrium spin polarization of a two-dimensional hole gas in an InGaAs QW due to the $p-d$ exchange interaction was reported in Ref.~\onlinecite{Aronzon-2009}. Also, the exchange fields in graphene layers coupled to yttrium iron garnet were used to achieve a strong modulation of spin currents~\cite{Kawakami-2017}.

\begin{figure*}[t]
\includegraphics[width=17cm]{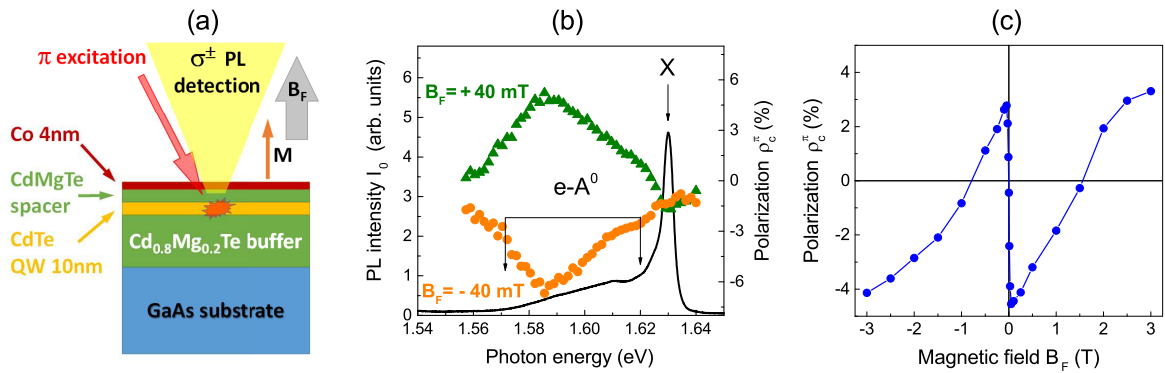}
\caption{(Color online) (a) Schematic presentation of the investigated structures and the PL excitation-detection geometry. Optical excitation is linearly ($\pi$) polarized. Circularly polarized $\sigma^+$ and $\sigma^-$ emission components are detected. The thickness of the Cd$_{0.8}$Mg$_{0.2}$Te buffer is 3~$\mu$m, the QW width is 10~nm, the Co thickness $d_{\rm Co} \approx 4$~nm and the spacer thickness is in the range of $d_S =5-10$~nm. In an external magnetic field $B_F \geq 50$~mT the interfacial FM magnetization $\mathbf{M}$ is directed perpendicular to the sample surface. (b) Spectra of PL intensity (black line) and degree of circular polarization (colored symbols). Excitation photon energy $\hbar\omega_{\rm exc}=1.7$~eV. (c) Magnetic field dependence of FM induced circular polarization $\rho_c^\pi(B_F)$. The polarization is averaged over the spectral range of the $e-A^0$ PL band ($1.57-1.62$~eV). All measurements are performed at $T_{\rm bath}=2$~K.}
\label{fig:PL-plus}
\end{figure*}

Recently, a new type of proximity effect was observed in a hybrid structure composed of a few nanometer thick Co layer which is deposited on top of a CdTe/CdMgTe semiconductor QW structure. The proximity effect was manifested in a FM induced spin polarization of holes bound to shallow acceptors in the QW~\cite{NP-Korenev-2016}. The polarization of the holes takes place due to an {\it effective $p-d$ exchange interaction} between the FM ($d$-system) and the QW holes ($p$-system). In this case, the FM produces an effective magnetic field which acts on the acceptor-hole spins and consequently leads to an equilibrium spin polarization of the holes. The main feature of this indirect exchange interaction is its long-range character, i.e. the proximity effect is almost constant with increasing thickness of the CdMgTe spacer between the FM and the QW layers up to 30~nm. This length scale is significantly larger than the 1-2~nm distance required for a significant overlap of wavefunctions in the direct exchange interaction between the QW holes and the magnetic ions in the FM. In Ref.~\onlinecite{NP-Korenev-2016} it was conjectured that the long-range indirect exchange originates from {\it exchange} of elliptically polarized acoustic phonons which exist in the FM layer close to the magnon-phonon resonance~\cite{Kittel-1958} and can penetrate into the SC layer. This mechanism was used later to explain the influence of elliptically polarized phonons on the magnetic properties of materials~\cite{Cavaleri-2017}. However, the spin polarization of the acceptors and the resulting circular polarization of the photoluminescence (PL) depend not only on the exchange splitting between the spin levels of the holes $\Delta_{pd}$ but also on other factors such as the temperature, the ratio of lifetime and spin relaxation time of the holes etc. Therefore, the polarization of the PL evaluated in Ref.~\onlinecite{NP-Korenev-2016} can be considered only as rough estimate for the splitting $\Delta_{pd} \approx 50~\mu$eV, and it is necessary to perform a direct measurement of the spin splitting of the acceptor holes using complementary techniques.

In this paper, we report on the investigation of the FM induced spin splitting of the acceptor bound holes in a CdTe QW located in close proximity of a Co layer. While previous optical and electrical measurements were indirect requiring additional model assumptions for analysis, here we perform a direct measurement using spin-flip Raman scattering giving the dependence of the Stokes shift $\Delta_S$ on external magnetic field $B$. In strong magnetic fields, $\Delta_S(B)$ scales linearly with $B$. Extrapolation of these data to zero magnetic field reveals a finite offset of the Stokes shift due to the FM induced effective exchange field with a magnitude of $\Delta_{pd} = 50-100~\mu$eV. This offset varies only weakly on the CdMgTe spacer thickness also in ranges where wavefunction overlap is negligible so that it has to be attributed to a long-range $p-d$ interaction. In addition, we show that the $s-d$ exchange interaction between conduction band electrons and the FM as well as the corresponding $p-d$ contribution for free valence band holes are negligible. These results are surprising from the viewpoint of standard theory of exchange interaction which is proportional to the overlap of the wavefunctions of the interacting particles. However, they are in line with the conjecture of an indirect exchange mediated by elliptically polarized phonons in FM-SC hybrid structures~\cite{NP-Korenev-2016} and therefore corroborate this model.

The paper is organized as follows. First, in Sec.~II we describe the proximity effect based on PL data recorded in a wide range of magnetic fields up to 3~T. Next, we present the results on spin-flip Raman scattering in Sec.~III. In Sec.~IV time-resolved data on pump-probe Kerr rotation are given where we evaluate the influence of the FM on the Larmor precession of the optically oriented holes and electrons. Finally, the results are discussed in Sec.~V.

\section{Ferromagnetic proximity effect}

The studied CdTe/Cd$_{0.8}$Mg$_{0.2}$Te QW structures were grown by molecular-beam epitaxy (MBE) on top of (100)-oriented GaAs substrates. The subsequent deposition of Co at room temperature was done without any intermediate contact to ambient atmosphere. Details on growth and characterization are given in Ref.~\onlinecite{NP-Korenev-2016}. A schematic presentation of the structure and of the geometry for PL measurements is shown in Fig.~\ref{fig:PL-plus}(a). The used gradient growth technique allowed variation of the thickness of both the Co layer and the Cd$_{0.8}$Mg$_{0.2}$Te spacer up to 10 and 30 nm, respectively. The 10~nm thick CdTe QW is sandwiched between Cd$_{0.8}$Mg$_{0.2}$Te barriers. The thickness of the Cd$_{0.8}$Mg$_{0.2}$Te buffer is about 3~$\mu$m. Most of studies are performed on  samples with a Co layer thickness of about 4~nm and a spacer thickness of $d_S=5-10$~nm. The samples are mounted in a split-coil helium bath cryostat with a variable temperature insert. The magnetic field is applied in the Faraday geometry parallel to the structure growth axis ($\mathbf{B}\|z$). In the PL measurements, excitation of electron-hole pairs in the QW layer is accomplished by picosecond optical pulses emitted by a tunable Ti:Sapphire laser at a repetition frequency of 75.75~MHz. The photon energy $\hbar\omega_{\rm exc}$ is kept below the band gap energy of the Cd$_{0.8}$Mg$_{0.2}$Te barriers ($\sim 1.9$~eV) in order to generate carriers in the QW layer only. The emission is analyzed and detected by a spectrometer equipped with a charge-coupled-device camera and a streak camera for time-integrated and time-resolved measurements, respectively.

Figure~\ref{fig:PL-plus} summarizes the time-integrated data on the ferromagnetic proximity effect. These PL data are  measured in the Faraday geometry on the sample with $d_S=10$~nm at a bath temperature of $T_{\rm bath}=2$~K. The total PL intensity $I_0=I^\pi_++I^\pi_-$ and degree of circular polarization $\rho^\pi_c$ spectra are shown in Fig.~\ref{fig:PL-plus}(b). The degree of circular polarization is defined as $\rho_c^\pi = (I^\pi_+-I^\pi_-)/(I^\pi_++I^\pi_-)$, where $I^\pi_+$ and $I^\pi_-$ are the $\sigma^+$- and $\sigma^-$-polarized emission intensities of the PL under linear polarized excitation, as indicated with the $\pi$ in the superscript. Already in weak magnetic fields $B_F=\pm40$~mT a circular polarization of several percent appears in the spectral range of the low energy PL band from $1.57-1.62$~eV, which corresponds to recombination of conduction band electrons with holes bound to acceptors (the $e-A^0$ band). This effect was studied in detail in our previous work where we demonstrated that:\cite{NP-Korenev-2016} (i) the circular polarization appears due to a FM induced spin polarization of the acceptor bound holes; (ii) the effect is induced by an {\it interfacial} FM which is formed at the Co/CdMgTe interface with a magnetization $\mathbf{M} || z$ and an out-of-plane (perpendicular) anisotropy (see Fig.~\ref{fig:PL-plus}). In weak magnetic fields the magnetization of the Co layer $\mathbf{M_{\rm Co}}$ is located in the plane of the structure ($\mathbf{M_{\rm Co}} \perp z$) and does not contribute to the circular polarization of the PL.

\begin{figure}[htb]
\includegraphics[width=0.9\columnwidth]{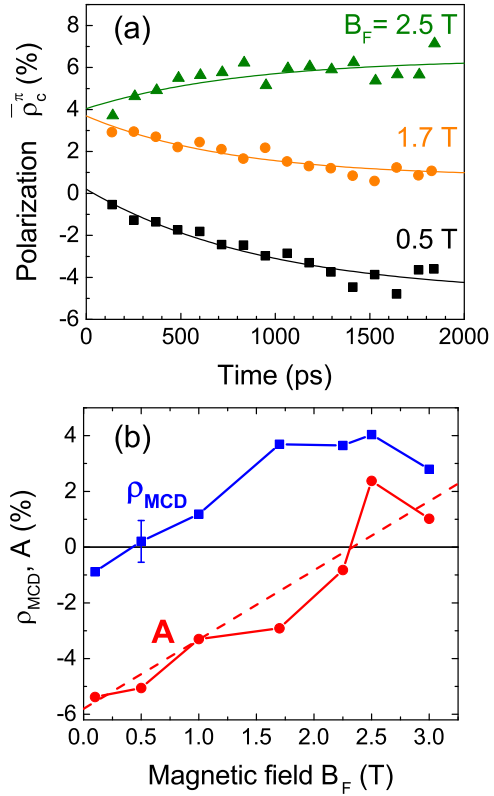}
\caption{(Color online) (a) Transients of circular polarization $\bar{\rho}_c^\pi(t)$ for different magnetic fields $|B_F|=$~0.5, 1.7, 2.5~T. Solid lines are fits to the data with Eq.~\eqref{eq:MCD}. (b) Magnetic field dependence of $\rho_{\rm MCD}$ and $A$. Dashed line is a fit to the data for $A$ using Eq.~\eqref{eq:PL-Delta} with $\Delta_{pd}=50\pm10~\mu$eV, $|g_A|=0.4\pm0.1$ and $T=5$~K.}
\label{Fig:PL-TRPL}
\end{figure}

Here, we extend the measurements of $\rho_c^\pi(B_F)$ to a larger magnetic field range up to 3~T, where an out-of-plane magnetization of the Co FM layer is present. Figure~\ref{fig:PL-plus}(c) shows the FM induced dependence $\rho_c^\pi(B_F)$ averaged across the spectral range from $1.57 - 1.62$~eV as function of the magnetic field $B_F$ in the Faraday configuration. In strong fields $B_F>0.25$~T the polarization increases with $B_F$ and changes its slope to a weaker dependence around 2~T, which is close to the saturation field of Co $4\pi M_{\rm Co}=1.7$~T~(see Ref.~\onlinecite{Chikazumi}). At first glance this behaviour could be attributed to a spin polarization of the holes due to exchange interaction with the Co where the exchange constant $J_{pd}$ has the opposite sign as that of the interfacial FM. However, care should be exercised here because there is a significant contribution of magnetic circular dichroism (MCD) to the data as follows from time-resolved photoluminescence (TRPL) measurements.

Figure~\ref{Fig:PL-TRPL}(a) shows transients of the antisymmetric term of the polarization degree $\bar{\rho}_c^\pi(|B_F|) = [\rho_c^\pi(+B_F) - \rho_c^\pi(-B_F)]/2$. The data can be well described with the following expression
\begin{equation}
\label{eq:MCD}
\bar{\rho}_c^\pi(t) = \rho_{\rm MCD} + A(1-e^{-t/\tau_S}),
\end{equation}
where the instantaneous polarization degree $\rho_{\rm MCD}$ results from the difference in absorption of $\sigma^+$ and $\sigma^-$ polarized light in the Co layer, the amplitude $A$ corresponds to the equilibrium polarization of the acceptor holes induced by the external magnetic field $B_F$ and the FM induced effective exchange field. $\tau_S$ is the spin relaxation time of polarized carriers, during which equilibrium populations of the spin levels are reached.

The magnetic field dependencies of $\rho_{\rm MCD}$ and $A$ evaluated from fits to the $\bar{\rho}_c^\pi(t)$ transients are shown in Fig.~\ref{Fig:PL-TRPL}(b). Obviously the MCD saturates at $B_F \approx 1.7$~T, while the amplitude $A$ continuously grows with $B_F$. The increase of $A$ with magnetic field is related to an additional equilibrium polarization of the holes due to thermalization between the spin levels. For small splittings ($A \ll 1$) the field dispersion of $A$ can be approximated by
\begin{equation}
\label{eq:PL-Delta}
A = \frac{\mu_B |g_A| B - \Delta_{pd}}{2 k_B T}
\end{equation}
where $\mu_B>0$ is the Bohr magneton, $k_B$ is the Boltzmann constant, and $g_A$ is the Land\'e factor of the acceptor which determines the splitting of the heavy hole states with angular momentum projections $J_z=\pm3/2$ onto the quantization axis of the QW. Since the amplitude $A$ does not saturate in magnetic fields $B_F>1.7$~T (see Fig.~\ref{Fig:PL-TRPL}(b)), we conclude that the contribution of the Co film to the proximity effect is negligible. Using Eq.~\eqref{eq:PL-Delta} we obtain $\Delta_{pd}=50\pm10~\mu$eV and $|g_A|=0.4\pm0.1$ (see the dashed line in Fig.~\ref{Fig:PL-TRPL}(b)). This evaluation depends, however, sensitively on the actual temperature of the crystal lattice $T$ in the illumination area which we assumed to be $T=5$~K, i.e. about 3~K higher than the bath temperature of $T_{\rm bath}=2$~K. Laser heating of the crystal lattice due to optical excitation is in agreement with our previous studies on optical orientation of Mn ions in GaAs~\cite{Akimov-2011}.

Thus, TRPL polarization measurements as applied up to now can be used to estimate the exchange energy splitting $\Delta_{pd}$ but this requires an accurate knowledge of $T$. Such a precise assessment is, however, hardly possible, but every determination of the crystal temperature is subject of considerable inaccuracies. In the following, we therefore present other methods that can be used for a direct measurement of the exchange energy which does not require any estimates.

\section{Evaluation of ${\bf p-d}$ exchange interaction via spin-flip Raman scattering}

\begin{figure}[htb]
\includegraphics[width=\columnwidth]{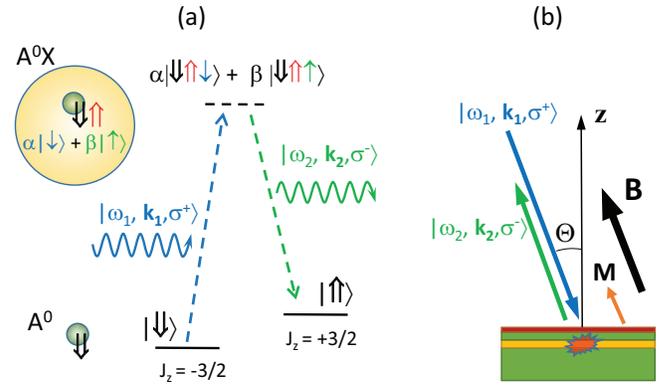}
\caption{(Color online) Schematic presentation of (a) SFRS for acceptor bound heavy hole; (b) geometry of SFRS experiment with $\Theta=20^\circ$. Black bold arrows $\Uparrow$ and $\Downarrow$ correspond to $z$-projections of angular momentum of acceptor hole, $J_z$, equal to $+3/2$ and $-3/2$, respectively. Red bold arrow $\Uparrow$ corresponds to $z$-projection of angular momentum of heavy hole in exciton which is equal to +3/2. Thin arrows $\uparrow$ and $\downarrow$ correspond to electron spin projection on $z$ axis, $+1/2$ and $-1/2$, respectively. Coefficients $\alpha$ and $\beta$ determine the mixing of electron spin states in external magnetic field and depend on angle $\Theta$.}
\label{Fig:SFRS-schema}
\end{figure}

Resonant spin-flip Raman scattering (SFRS) allows measuring the magnetic field induced splitting of the spin levels of charge carriers in semiconductor QW structures~\cite{SapegaPRB94,Akimov-2011,Sirenko-1997,Debus2013}; moreover, it can be also exploited to evaluate exchange energies by which different spin configurations are separated~\cite{Debus2014}. As we will demonstrate in the following, in contrast to polarization-resolved PL measurements, SFRS grants access to the effective $p-d$ exchange constant in the hybrid structures studied here. The physics of SFRS for a hole bound to an acceptor is shown in Fig.~\ref{Fig:SFRS-schema}.

Initially, the exciting photon in state $|\omega_1, \mathbf{k_1},\sigma^+ \rangle$ with optical frequency $\omega_1$ and circular polarization $\sigma^+$ propagates along the magnetic field direction $\mathbf{k_1}\|\mathbf{B}$. The $|\pm3/2\rangle$ ground states of the heavy hole bound to an acceptor $A^0$ in the QW are the eigenstates of the angular momentum projection onto the direction $z$ perpendicular to the QW plane, $J_z=\pm3/2$ (black bold arrows $\Uparrow$ and $\Downarrow$ in Fig.~\ref{Fig:SFRS-schema}(a)). In the absence of $p-d$ exchange interaction, the Zeeman splitting of the spin levels is given by $E_{\pm3/2}=\pm\frac{1}{2}\mu_Bg_AB$. In the intermediate SFRS state the $A^0X$ complex given by an exciton bound to a neutral acceptor is created. For $\sigma^+$ excitation,  the angular momentum projection of the heavy hole in the exciton is equal to +3/2 (red bold arrow $\Uparrow$ in Fig.~\ref{Fig:SFRS-schema}(a)), while the spin of the acceptor bound hole is equal to $J_z=-3/2$ (see Fig.~\ref{Fig:SFRS-schema}(a))\cite{SapegaPRB94}. The exchange interaction between the exciton heavy hole and the acceptor bound hole can lead to a mutual flip of their spins with conservation of total angular momentum. In the next step, the exciton is annihilated and a photon is emitted with optical frequency $\omega_2$ and opposite circular polarization $\sigma^-$. Here, energy conservation is fulfilled only for the initial and final states (photon and acceptor), but not in the intermediate state (exciton bound to neutral acceptor). In the final state we obtain the emitted photon $|\omega_2, \mathbf{k_2},\sigma^-\rangle$ and the acceptor with $J_z=+3/2$. Thus, the energy of the emitted photon is $\hbar\omega_2 = \hbar\omega_1 - \mu_B |g_A| B$, which is shifted into the Stokes region.

\begin{figure*}[htb]
\includegraphics[width=15cm]{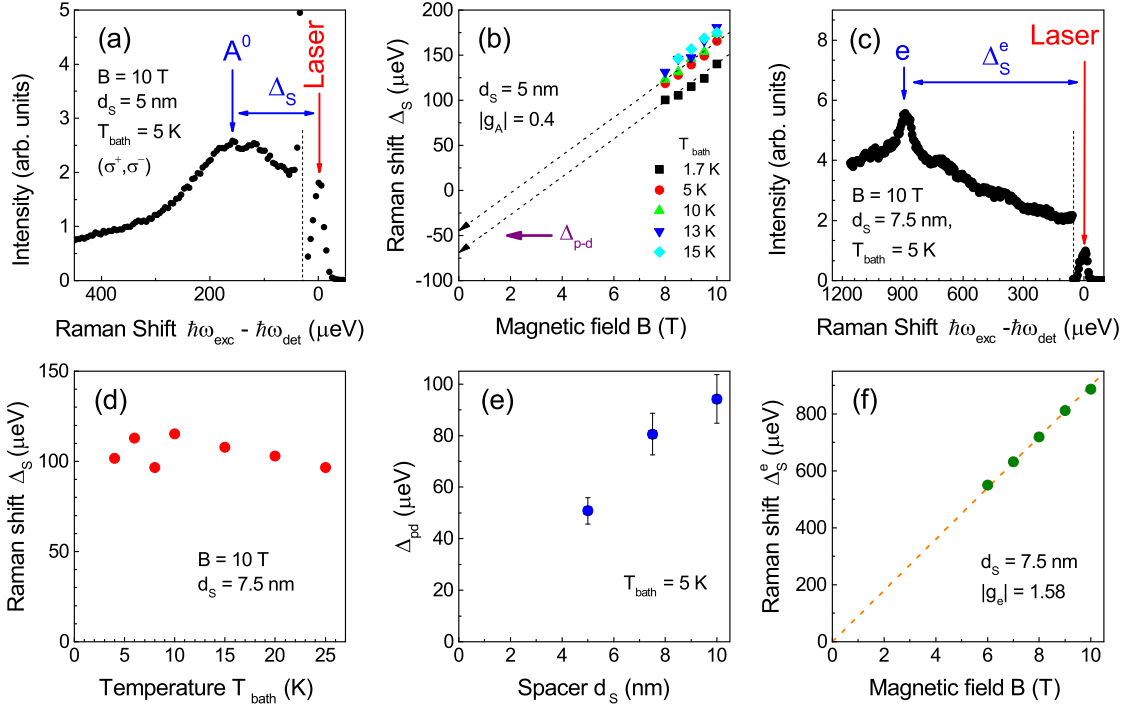}
\caption{(Color online) (a) Example of SFRS for excitation energy $\hbar\omega_{\rm exc}=1.610$~eV measured in cross-polarized excitation/detection ($\sigma^+,\sigma^-$) configuration. $B=10$~T, $T_{\rm bath}=5$~K and $d_S=5$~nm.  Vertical arrow indicates peak position of the hole spin flip at the acceptor. Vertical dashed lines in panels (a) and (c) correspond to laser cut-off energies, where a filter with $0.01\%$ transmission was introduced in the detection path in order to accurately measure the spectral position of laser line. (b) Magnetic field dependence of Raman Stokes shift $\Delta_S$ of hole spin flip line for various bath temperatures $T_{\rm bath}$ (symbols). Dashed arrows correspond to linear fits using Eq.~\eqref{eq:Exchange} with $|g_A| =0.4$ and point to the energy offset $\Delta_{pd}$.  (c) Example of SFRS for $\hbar\omega_{\rm exc}=1.615$~eV showing electron spin flip line. $B=10$~T, $T_{\rm bath}=5$~K and $d_S=7.5$~nm.  (d) Temperature dependence of Raman shift $\Delta_S$ at $B=10$~T for the structure with $d_S=7.5$~nm. (e) Exchange energy $\Delta_{pd}$ for various spacer thicknesses $d_S$. $T_{\rm bath}=5$~K. (f) Magnetic field dependence of Raman shift for electron spin flip $\Delta_S^e$. Dashed line is linear fit to data with $|g_e|=1.58$. All SFRS data are measured for $\Theta=20^\circ$.}
\label{Fig:SFRS-shift}
\end{figure*}

In Faraday geometry ($\mathbf{B}\|z$) the transition described above is forbidden because the angular momentum of the hole in the $A^0X$ complex should change by three quanta, $\Delta J_z =3$, while the angular momentum of the photon $\Delta l$ in the back-scattering geometry ($\mathbf{k_1}=-\mathbf{k_{2}}$) changes by 0 or $\pm2$ only. For the observation of an SFRS line corresponding to the transition of the hole between its Zeeman levels we use therefore an oblique field geometry, namely an angle $\Theta$ between the $z$-axis and magnetic field $B$ of $20^\circ$ is chosen (see Fig.~\ref{Fig:SFRS-schema}(b)). In this geometry, the magnetic field induces a mixing of the electron states with spin projections $+1/2$ and $-1/2$ along $z$ (thin arrows $\uparrow$ and $\downarrow$ in Fig.~\ref{Fig:SFRS-schema}(a)) which allows for observing SFRS in crossed circular polarizations~\cite{SapegaPRB94}. For an efficient SFRS process, it is necessary to tune the laser photon energy into resonance with the $A^0X$ transition (1.610~eV). In case of a noticeable $p-d$ exchange interaction between the magnetic ions in the FM layer ($d$-system) and the holes bound to acceptors in the QW ($p$-system) the splitting $\Delta_S(B)$ of the $A^0$ states is determined not only by the external magnetic field $B$, but also by the additional contribution due to the effective exchange field from the FM. Therefore, the resulting splitting is given by
\begin{equation}
\label{eq:Exchange}
\Delta_S(B)=\mu_B |g_A| B - \Delta_{pd}m_z,
\end{equation}
where $m_z$ is the $z$-projection of the unit vector $\mathbf m$ along the magnetization $\mathbf M$. Eq.~\eqref{eq:Exchange} is valid for large magnetic fields, when the first term on the right hand side is larger than the second one, i.e. $\Delta_S(B)>0$. Here, we use the fact that the $p-d$ exchange interaction between the magnetic ions and the heavy holes in a QW structure is strongly anisotropic, i.e., it is described by the Ising Hamiltonian $\frac{1}{3}\Delta_{pd}m_zJ_z$ ~\cite{Merkulov-1995}. In strong magnetic fields, the FM is fully magnetized along the $B$-direction and the dependence $\Delta_S(B)$ is a straight line with an offset given by the exchange constant $\Delta_{pd}$.  For small $\Theta$, the projection $m_z=\cos\Theta\approx1$ and $|g_A|$ corresponds to the longitudinal acceptor $g$ factor, which determines the Zeeman splitting for $B$ applied along the $z$-direction. In our case $\Theta=20^\circ$ which allows one to use $\cos\Theta = 1$ in the evaluation of the exchange energy $\Delta_{pd}$ with an accuracy of 7\%.

Figure~\ref{Fig:SFRS-shift} summarizes the data on the SFRS corresponding to the spin flip of the electron ($e$) and the hole bound to an acceptor ($A^0$). For $B=10$~T under resonant excitation of the $A^0X$ transition with photon energy $\hbar\omega_{\rm exc}= 1.610$~eV, the spin flip of the acceptor bound hole is observed for crossed orientations of polarizer and analyzer ($\sigma^+,\sigma^-$) as shown in Fig.~\ref{Fig:SFRS-shift}(a). The signal is given by the broad line with a Raman shift of $\Delta_S =160~\mu$eV close to the laser line. Figure~\ref{Fig:SFRS-shift}(b) shows the magnetic field dependences of the Raman shift of the acceptor bound hole $\Delta_S$ for various temperatures $T_{\rm bath}$. The data are well described by Eq.~\eqref{eq:Exchange} with the hole g factor $|g_A|=0.4$ which determines the slope of the line. The offset $\Delta_{pd}\approx 50~\mu$eV for $T_{\rm bath}=5$~K and depends weakly on temperature. A weak dependence of $\Delta_{pd}$ on $T_{\rm bath}$ follows also from Fig.~\ref{Fig:SFRS-shift}(d), where the temperature dependence of the Raman shift $\Delta_{S}$ for a fixed magnetic field $B=10$~T is shown. Such behavior cannot be attributed to an exchange interaction with paramagnetic ions or FM Co clusters diffused into the QW during the growth process. The magnetization of ions should decrease strongly with increasing temperature from 2 to 25~K, which is in contrast to our observations (see Fig.~\ref{Fig:SFRS-shift}(b) and (d)). Thus, we conclude that the SFRS demonstrates the splitting of the acceptor bound hole in the FM induced exchange field. The striking feature of this interaction is its long range nature. Figure~\ref{Fig:SFRS-shift}(e) shows the splitting $\Delta_{pd}$ ${\sl vs}$ the spacer thickness evaluated from magnetic field dependences of $\Delta_S(B)$ measured on corresponding samples. We observe a splitting of about 100~$\mu$eV even for spacers as large as 10~nm. This distance is significantly larger than the penetration depth of electron and hole wavefunctions of maximum 1-2~nm into a FM layer that would be required to obtain a considerable direct exchange interaction~\cite{NP-Korenev-2016}.

\begin{figure}[htb]
\includegraphics[width=0.7\columnwidth]{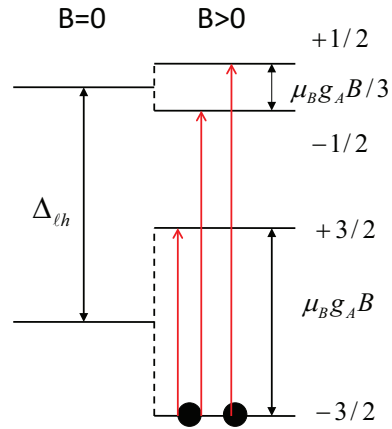}
\caption{(Color online) Scheme of optical transitions involved in SFRS of acceptor bound hole (red arrows). Lower energy doublet with angular momentum projections $\pm 3/2$ corresponds to heavy hole states where splitting in magnetic field is depicted for the case of $g_A>0$. Upper energy doublet with angular momentum projections $\pm1/2$ corresponds to light hole states. $\Delta_{lh}$ is energy splitting between heavy and light holes bound to acceptor.}
\label{Fig:SFRS-acceptor}
\end{figure}

The offset in the magnetic field dependence of the acceptor bound hole Raman shift $\Delta_S(B)$ has to be considered with considerable care. Apart from the FM induced exchange field, the offset may result from the energy splitting between the heavy and light holes bound to an acceptor. The magnitude of this splitting is about $\Delta_{lh}\approx 1$~meV~\cite{SapegaPRB94}. For the magnetic fields $B \leq 10$~T used in our experiments, the Zeeman splitting of the hole states $\mu_B |g_A| B$ is clearly less than $\Delta_{lh}$, which results in the transition scheme shown in Fig.~\ref{Fig:SFRS-acceptor}. At low temperatures, the lowest energy heavy hole state with angular momentum projection $J_z=-3/2$ is populated. From this state, there are three possible transitions which are shown with the red arrows in Fig.~\ref{Fig:SFRS-acceptor}. It follows that a decrease of the magnetic field leads to vanishing of the $|-3/2\rangle \rightarrow |+3/2\rangle$ spin flip transition energy. However, the transitions $|-3/2\rangle \rightarrow |-1/2\rangle$ and $|-3/2\rangle \rightarrow |+1/2\rangle$ have a positive offset corresponding to $\Delta_{lh}$. We emphasize that our results cannot be attributed to such behaviour because: (i) the offset in Fig.~\ref{Fig:SFRS-shift}(b) is negative and (ii) the magnitude of exchange energy $\Delta_{pd} < 100~\mu$eV is significantly smaller than $\Delta_{lh}$. Moreover, the magnetic field dependence of $\Delta_S(B)$ in CdTe QW structures without Co layer shows linear behavior which approaches zero when extrapolated to zero field, i.e. no offset is detected in this case.

Therefore, the observation of SFRS on the acceptor bound hole corresponds to the spin-flip transition $|-3/2\rangle \rightarrow |+3/2\rangle$ and the offset is related to the heavy hole splitting in the effective exchange field from the FM. Transitions to the light hole states with $J_z=\pm1/2$ were not detected in the investigated samples which may be attributed to spectral broadening of the Raman line due to fluctuations of $\Delta_{lh}$.

The SFRS signal related to the heavy hole spin flip disappears when the exciting laser photon energy is increased and approaches the exciton resonance $X$ (see the PL spectrum in Fig.~\ref{fig:PL-plus}(b)). In this case the spin flip of the electron dominates the SFRS spectrum, which is shown in Fig.~\ref{Fig:SFRS-shift}(c) for $\hbar\omega_{\rm exc}=1.615$~eV. Figure~\ref{Fig:SFRS-shift}(f) presents the magnetic field dependence of the Stokes shift for the electron spin-flip $\Delta_S^e(B)$. The shift follows a linear dependence with the electron $g$ factor $|g_e|=1.58$ and does not show any measurable offset~\cite{Sirenko-1997}. This indicates that the effective $s-d$ interaction between the conduction band electrons in the QW and the FM layer is negligibly small as compared with the $p-d$ interaction of the QW heavy holes.

We also note that we do not observe a SFRS signal related to the free heavy hole which is not bound to the acceptor. Its absence may be due to strong fluctuations of the free hole $g$ factor leading to a significant broadening of the SFRS line. For detecting the spin splitting of the unbound heavy hole $\hbar\Omega_h$, we use a transient pump-probe technique as described below.

\section{Larmor spin precession of valence band holes}

Transient pump-probe Kerr rotation in the vicinity of the exciton resonance allows us to measure the frequency of the Larmor precession of electrons $\Omega_e$ and holes $\Omega_h$ in CdTe/(Cd,Mg)Te QWs~\cite{Zhukov-2007}. Thereby circularly polarized pump pulses photoexcite carriers with optically oriented spin polarization parallel to the growth direction ($z$-axis). In a transverse magnetic field $\mathbf{B}\|x$ the subsequent spin precession leads to transient oscillations of the $z$-component, $S_z$, of the spin polarization which is detected by the Kerr rotation of the linearly polarized probe beam when the delay between the pump and probe pulses $t$ is varied. The electron and hole spins precess with different Larmor precession frequencies due to the difference in their $g$ factors. The electron $g$ factor in CdTe QW is close to isotropic, while the heavy hole one has a strong anisotropy.

Our experiment requires an oblique magnetic field since the $z$-component of magnetic field has to induce a magnetization of the FM layer, while the $x$-component is required to observe the oscillatory precession signal. We stress that the pump-probe signal is observed in the studied FM-SC hybrid structures only when the excitation photon energy is tuned to the QW exciton resonance. This indicates that the experimental data monitor the spin dynamics of photoexcited carriers in the QW and not in the FM. Moreover, we get exclusively access to the Larmor precession of the conduction band electrons and valence band holes because an efficient optical orientation of the photoexcited carriers occurs only for resonant excitation of the excitons, whose oscillator strength is significantly larger than that of  the excitons bound to acceptors.

\begin{figure}[htb]
\includegraphics[width=\columnwidth]{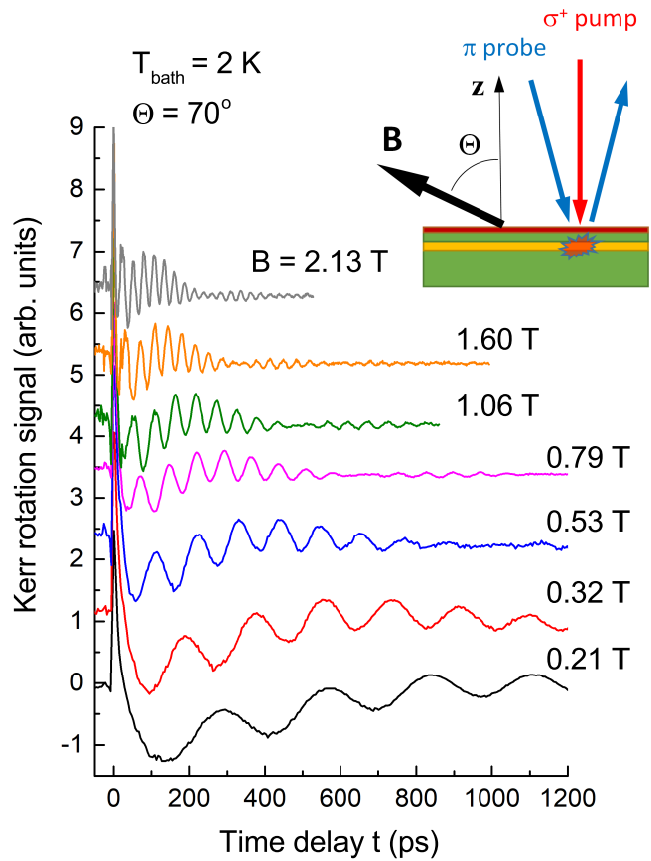}
\caption{(Color online) Transient pump-probe Kerr rotation signal measured as function of pump-probe delay for various magnetic fields. Photon energy of pump and probe $\hbar\omega_p=1.627$~eV is tuned in resonance with exciton transition. $T_{\rm bath}=2$~K,  $d_S=10$~nm and $\Theta=70^\circ$. Inset shows schematically geometry of experiment.}
\label{Fig:PP-transients}
\end{figure}

\begin{figure}[htb]
\includegraphics[width=0.8\columnwidth]{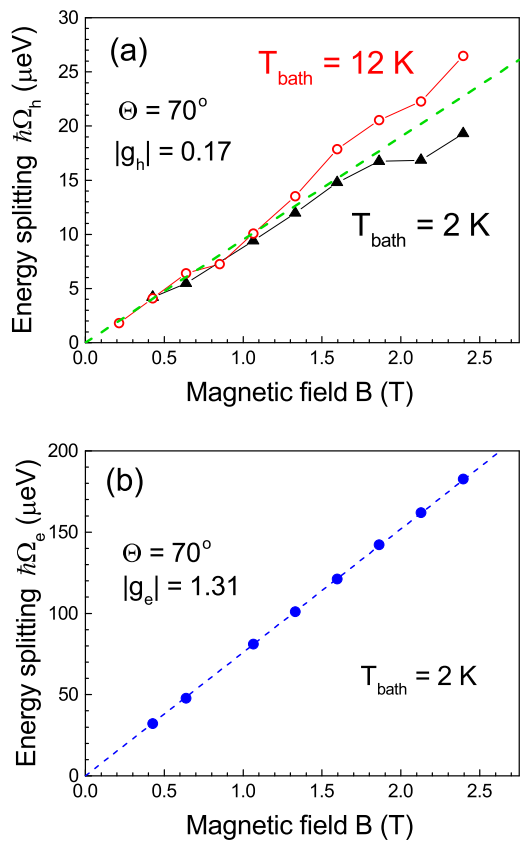}
\caption{(Color online) Magnetic field dependence of Zeeman splitting for (a) holes $\hbar\Omega_h$ for two temperatures of 2~K and 12~K and (b) electrons $\hbar\Omega_e$ evaluated from pump-probe transients. Dashed lines are linear fits to the data with $|g_h|=0.17$ in (a) and $|g_e|=1.31$ in (b). $d_S=10$~nm and $\Theta=70^\circ$.}
\label{Fig:PP-Freq}
\end{figure}

Figure \ref{Fig:PP-transients} shows corresponding transient Kerr rotation signals in different magnetic fields. The inset shows schematically the geometry of the experiment where the magnetic field is tilted by an angle $\Theta=70^\circ$ with respect to the $z$-axis. The transient signals comprise two contributions. The first one corresponds to a signal with high oscillation frequency and is attributed to the electron spin precession. The second contribution oscillates quite slowly and corresponds to the heavy hole spin dynamics with a small $g$ factor. Each of these oscillatory signals is well described with $A_i\cos(\Omega_it+\phi_i)$ which allows us to determine the magnetic field dependence of the Larmor precession frequencies $\Omega_i$ for the electrons ($i=e$) and holes ($i=h$). The data are summarized in Fig.~\ref{Fig:PP-Freq}.

For the holes, $\Omega_h(B)$ dependencies are shown for $\Theta=70^\circ$ at two different temperatures, 2~K and 12~K (Fig.~\ref{Fig:PP-Freq}(a)). At first glance the dependences appear to be linear across the whole range of magnetic fields with the corresponding $g$ factor $|g_h|=0.17$ which weakly depends on temperature. However, a closer look shows that $\Omega_h(B)$ shows small wiggles above about $B\approx1.5$~T. One possible explanation for the non-linear behavior of $\Omega_h(B)$ is the exchange interaction of the heavy holes with magnetic ions in the FM layer whose magnetization slowly varies with magnetic field. However, even if this effect is present its magnitude is expected to be rather small. Therefore, we conclude that the valence band holes are weakly coupled to the FM layer which is in contrast to the strongly interacting holes bound to acceptors as demonstrated in Section~III. The value of the heavy hole $g$ factor is determined from the relation $|g_h|= \sqrt{g_z^2\cos^2\Theta + g_x^2\sin^2\Theta}$. Taking $g_x\approx0$ we obtain $|g_z|\approx 0.5$ thereby.  This value is slightly larger than the $g$ factor of the acceptor bound hole $|g_A|=0.4$ extracted from the SFRS data, which indicates that indeed the pump-probe signal addresses the spin dynamics of unbound, free valence band holes.

The Larmor precession frequency of the electrons $\Omega_e(B)$ depends linearly on magnetic field (Fig.~\ref{Fig:PP-Freq}(b)), from which we evaluate the electron $g$ factor to be $|g_e|=1.31$. The slight difference between the values obtained from pump-probe and SFRS is related to the anisotropy of the electron $g$ factor~\cite{Sirenko-1997}. Note that the magnetic field dependence of $\Omega_e$ also does not show any offset.  Thus, the electrons do not experience a $s-d$ exchange interaction which is in accord with the SFRS data.

\section{Discussion}

The main result of our study is the direct measurement of the exchange energy $\Delta_{pd} = 50-100~\mu$eV for the effective $p-d$ interaction between the magnetic ions in the FM layer and the holes bound to acceptors in the semiconductor QW, without involving any model. This energy splitting of the hole spin levels is in agreement with our previous estimates in Ref.~\onlinecite{NP-Korenev-2016}, where $\Delta_{pd} \approx 50~\mu$eV was evaluated from polarization- and time-resolved PL measurements in weak longitudinal magnetic fields in which the interfacial FM layer resulted in a magnetization of the acceptor holes. Here, SFRS measurements have been performed in strong magnetic fields and, therefore, it is expected that an additional contribution from the Co layer to the exchange interaction is expected. This is because magnetic fields larger than 2~T are sufficient to saturate the out of plane magnetization of the Co film. However, in contrast to MCD the amplitude of the proximity effect $A(B)$ in Fig.~\ref{Fig:PL-TRPL}(b) increases linearly with magnetic field. Therefore, we conclude that the main contribution to the $p-d$ exchange interaction comes from the interfacial FM. The origin of the interfacial magnetic layer requires further studies. Currently, it is reasonable to assume that its formation is caused by chemical reaction of Co atoms with the Cd$_{0.8}$Mg$_{0.2}$Te material.

We observe no FM induced splitting of the spin levels of the valence band holes which are not bound to acceptors as well as of the conduction band electrons. The splitting of valence band holes has been evaluated from degenerate pump-probe Kerr rotation measurements under resonant excitation of excitons in the QW structure. This experiment differs significantly from SFRS which probes the acceptor bound holes under resonant excitation of excitons bound to neutral acceptors $A^0X$. Figure~\ref{Fig:PP-Freq}(a) demonstrates that the magnetic field dependence $\Omega_h(B)$ does not show a detectable offset and a deviation from a linear behavior. Thus, the pump-probe measurements clearly demonstrate that the exchange interaction between the valence band holes in the QW and the FM layer is small. The same result is obtained for the conduction band electrons where as well no offset in the magnetic field dependence of their Zeeman splitting is detected, both in SFRS and pump-probe.

A further result obtained from SFRS is that the exchange energy $\Delta_{pd}$ does not decrease with increasing spacer thickness for $d_S \le 10$~nm (see Fig.~\ref{Fig:SFRS-shift}(e)). This is in accord with our previous studies in Ref.~\onlinecite{NP-Korenev-2016}, where the suppression of PL intensity with decreasing $d_S$ gives a characteristic length of $1-2$~nm for the wavefunction penetration into the Co-layer. This distance is much smaller than the spacer range of $d_S=5-10$~nm addressed in the present study. Also, the FM induced polarization of the PL depends only weakly on $d_S =5-30$~nm~\cite{NP-Korenev-2016}. Therefore, we conclude that the effective $p-d$ exchange interaction between the Co ions in the FM and the holes bound to acceptors in the QW is not determined by their wavefunction overlap. These results are surprising from the viewpoint of the standard theory of exchange interaction whose strength is proportional to this overlap~\cite{Miotkowska-2001,Kossacki-Co-CdTe-2013}. Note, however, that this does not represent a contradiction because the exchange reported here is observed for holes bound to acceptors but is absent for conduction band electrons and valence band holes.

\begin{figure}[htb]
\includegraphics[width=\columnwidth]{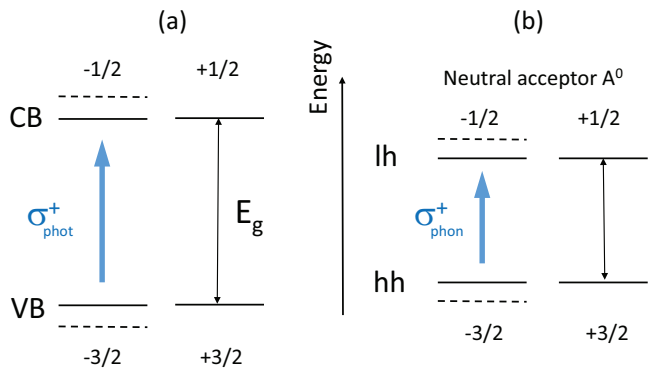}
\caption{(Color online) Optical (a) and phonon (b) {\it ac} Stark effect in semiconductors with $\sigma^+$ photons and phonons (blue arrows), respectively. In (a) the energy of photons is tuned below the band gap energy $E_g$ of semiconductor and results in a shift of electronic states in the conduction band (CB) with spin projection $S_z=-1/2$ and the valence band (VB) with angular momentum projection $J_z=-3/2$. In (b) phonons couple to transitions between between heavy (hh) and light (lh) hole states which are split by $\Delta_{lh}$. For $\sigma^+$ polarized phonons the energy shift occurs for the heavy hole state with $J_z=-3/2$ and the light hole state with $J_z=-1/2$. The $z$-axis is dictated by the photon or phonon propagation direction. The repulsion of energy levels is indicated with dashed lines. Both {\it ac} Stark effects result in a splitting of spin states and can be considered as generation of an effective {\it dc} magnetic field.}
\label{Fig:ac-Stark}
\end{figure}

In Ref.~\onlinecite{NP-Korenev-2016} we proposed that this kind of long-range interaction can be mediated by elliptically polarized acoustic phonons. The latter are strongly polarized in the vicinity of the magnon-phonon resonance in the FM~\cite{Kittel-1958}. In addition, phonons do not experience the electronic barrier between the QW and the FM layer. The characteristic frequencies of these elliptically polarized phonons (about 1~meV) are close to the energy splitting between the acceptor bound heavy $|\pm3/2\rangle$ and light $|\pm1/2\rangle$ holes  (quasi-resonant case) and significantly smaller than the corresponding splitting between the confined valence band states in the QW with 10~nm width. For example, if the phonons are mainly $\sigma^+$ polarized (with positive $z$-projection of angular momentum) the interaction with the holes couples the ground state $|-3/2\rangle$ with the excited state $|-1/2\rangle$ which consequently leads to an energy shift of the levels. This results in lifting of the Kramers degeneracy of the $|\pm3/2\rangle$ doublet in zero external magnetic field which is the phonon analog of the optical {\it ac} Stark effect and the inverse Faraday effect which occurs in case of illumination with elliptically polarized light in transparency region.

The optical Stark effect is a well established phenomenon in semiconductors \cite{Combescot-1989}. It takes place when an electromagnetic wave with $\sigma^+$ polarization couples the electronic states with angular momentum projection $-3/2$ in the valence band and $-1/2$ state in the conduction band as shown in Fig.~\ref{Fig:ac-Stark}(a). Due to the interaction with light these states experience an energy shift $\Delta \propto P^2/\delta$, where $\delta =  E_g - \hbar\omega$. Here, $\hbar\omega$ is the photon energy and $E_g$ is the energy gap of the semiconductor, $P$ is the dipole matrix element of the optical transitions between valence and conduction bands. For photons with $\hbar\omega < E_g$ repulsion between the electronic states takes place, i.e. $\Delta>0$. Similarly, in the case of the phonon Stark effect~\cite{NP-Korenev-2016}, a circularly polarized phonon couples the heavy (hh) and light (lh) hole acceptor states with angular momentum projections $-3/2$ and $-1/2$, respectively (see Fig. \ref{Fig:ac-Stark}(b)). The spin-phonon interaction for holes occurs due to the spin-orbit coupling of hole states in the valence band. In this case, the level shift is proportional to the square of the matrix element of the spin-phonon interaction divided by the detuning of the phonon frequency at the magnon-phonon resonance in the FM relative to the energy separation between the heavy and light hole acceptor levels in the QW.

In conclusion, our results are in agreement with the proposed model of an {\it effective $p-d$ exchange interaction} mediated by elliptically polarized phonons. Here the energy splitting of the acceptor bound holes has been measured directly and amounts to $\Delta_{pd} = 50-100~\mu$eV. This model explains the absence of a long-range $s-d$ exchange interaction because the spin-orbit interaction in the conduction band is much smaller than the one in the valence band.

\section{Acknowledgements}

We acknowledge the financial support by the Deutsche Forschungsgemeinschaft through the International Collaborative Research Centre 160. The partial financial support from the Russian Foundation for Basic
Research Grant No. 15-52-12017 NNIOa and the Russian Ministry of Education and Science (Contract No. 14.Z50.31.0021) is acknowledged as well. N.E.K. acknowledges support from RFBR (grant No. 15-52-12019). The research in Poland was partially supported by the National Science Centre (Poland) through Grants No. DEC- 2012/06/A/ST3/00247 and No. DEC-2014/14/M/ST3/00484, as well as by the Foundation for Polish Science through the IRA Programme co-financed by EU within SG OP.

\end{document}